# Multiphysics analysis and practical implementation of an ionic soft actuator-based microfluidic device toward the design of a POCT compatible active micromixer

Mohsen Annabestani, Sina Azizmohseni, Pouria Esmaeili-Dokht, Nahal Bagheri, Afarin Aghassizadeh and, Mahdi Fardmanesh*

*Abstract*—Electroactive-Polymers (EAPs) are one of the best soft materials with great applications in active microfluidics. Ionic ones (*i*-EAPs) have more promising features for being appropriate candidates to use in active microfluidic devices. Here, as a case study, we have designed and fabricated a microfluidic micromixer using an *i*-EAP named Ionic Polymer-Metal Composite (IPMC). In microfluidics, active devices have more functionality but due to their required facilities are less effective for Point of Care Tests (POCTs). In the direction of solving this paradox, we should use some active components that they need minimum facilities. IPMC can be one of these components, hence by integrating the IPMC actuator into a microfluidic channel, a micromixer chip was designed and put to the simulation and experimental tests. The result showed that the proposed micromixer is able to mix the micro fluids properly and IPMC actuator has adequate potential to be an active component for POCT-based microfluidic chips.

I. INTRODUCTION

Microfluidic systems or micro-total analysis systems (μ-TAS), are devices controlling liquids in geometrically constrained channels. Liquids behave distinctively in microfluidic channels from that typically anticipated due to the small size of these channels. Surface tension, fluidic resistance, and energy dissipation might be contributing factors behind this difference. This implies that we need an alternate hypothesis, technology, and innovation for managing these behaviors. In science, microfluidics is a field that assumes the main job in this issue, and it can support scientists and researchers to discover practical solutions for the inquiry that "how can we use and manipulate the fluids in such small volumes [1]". Microfluidics has different applications in various fields such as medicine [2], diagnostics especially Point Of Care (POC) tests[3], chemical analysis [4], electronic industries [5], Lab-on-a-chip (LOC)[6], Organ-on-a-chip (OOC) [7], etc. Microfluidic devices are generally portrayed in two principal sorts of "active" and "passive" devices. Active devices are the ones that need at least one outside actuation to make an action. For instance, particle sorting issues for the most part profit by applying di-electrophoresis [8], magnetic [9], optical [10], acoustic [11], or any other stimulation for arranging the particles.

Nevertheless, the essential methodology in passive devices is attempting to work dependent on making a few interactions between the device's various segments, such as small-scale channels structure and their geometry, the flow field, and so on. In spite of the fact that there is an assortment of utilizations that use passive chips, some of the time a microfluidic system can utilize some active operation (e.g., active sorting) or active elements (such as valves, pumps, and mixers) in order to productively control the liquids [12]. Active valves, pumps, and mixers use various actuators. Known actuators include thermopneumatic, pneumatic, piezoelectric, hydraulic, electrostatic displacements, electromagnetic actuators, or Electroactive-polymers (EAPs) [12]. A large portion of the ordinary active components utilized in a microfluidic device are challenging to control and assemble. Moreover, they need extraordinary facilities, and they do not seem reasonably affordable for everybody. The above-mentioned limitations are against a large number of the several aims of microfluidic devices including availability, the development of POCs, at-home services, low economic costs, etc. However, ionic EAPs (*i*-EAPs) are not like typical actuators since they do not have the limitations referenced previously. *i*-EAPs can be scaled down in small sizes and coordinated into MEMSs and μ-TASs. Furthermore, they are electrically driven and require extremely low input voltage. This scope of input voltage and power is minimal cost, which is omnipresent wherever by a PC, a simple telephone charger, a smartphone, or even a battery. The referenced preferences of *i*-EAPs guarantee us that they are generally excellent possibility for active components and offer a safe-ground to utilize them in active microfluidic devices. Hence, *i*-EAPs can offer some promising solutions for some practical application of active microfluidics.

We can categorize *i*-EAPs in three main categories, Ionic carbon nanotube-based actuators (*i*-CNTAs), Conducting or sometimes conjugated polymer actuators (CPAs) and Ionic polymer-metal composites (IPMCs). Because of their especial qualities, *i*-EAPs can be appropriate candidates as active components of microfluidic devices, particularly microvalves, micropumps, and micromixers. Between pumps, valves, and mixers, we can find several *i*-EAP based microfluidic micropumps [13, 14] but microvalves and especially micromixers have less used *i*-EAPs to work actively. In this paper, it will be simulated and practically shown that the IPMC actuators have adequate potential to be active components of microfluidic micromixers.

The rest of this paper is described in six parts. In part II, the IPMC actuator will be introduced as proposed candidate *i*-EAP that can likely function as an active component of active microfluidic micromixer. The proposed IPMC-based micromixer will be described in part III. The reasoning behind the purposed approach is described thoroughly in section IV, and related equations are presented. The 3D model for COMSOL simulation and results are shown in part V. Finally, in part VI, the fabrication process of the purposed micromixer is explained, and experimental results are discussed.

## II. IONIC POLYMER-METAL COMPOSITE (IPMC)

One group of soft *i*-EAPs are IPMCs, which can be used as actuator or a sensor[15, 16]. This group has been widely used in biomedical applications due to their promising highlights. Lightness, high toughness, and low density are some of the above-mentioned promising highlights. Moreover, showing high stimulus strain with a low driving voltage are another important advantage [17, 18]. On the other hand, elimination operations should be necessarily done on several drawbacks of this smart material. For instance, a portion of these downsides include not being strong enough in several applications, some undesirable impacts in its performances, such as back relaxation effects [19, 20]. A cantilever beam is a common structure for an IPMC, consisting of a thin membrane (usually Nafion) that is coated by a metallic plate on each side (Pt in most cases). The internal ionic content of an IPMC (hydrated sodium cations in most applications) moves toward the cathode electrode once the voltage is applied between the two plates of an IPMC through its diameter. The transportation of the ionic substance makes the Nafion swell around the cathode side and makes a bending response towards the anode side as appeared in Fig.1. On the other hand, a low voltage is generated between two electrodes across the membrane, when we deform or bend the IPMC. The amplitude of generated voltage is proportional to the applied deformation [16, 17]. It can be concluded that an IPMC can function both as a sensor and as an actuator. The IPMC's ability to be scaled down as well as its promising possibilities has made it an appropriate candidate for MEMS and active microfluidic devices. The fundamental focal point of IPMC-based active microfluidic devices is their applications with micropumps actuated with IPMCs [13]. Microvalves [12] and IPMC-based micromixers [21] are Other applications of IPMCs as an active component in microfluidics. The primary goal of this paper is to utilize IPMC as an actuator in a micromixer so as to mix the laminar liquids with low Reynolds number. The proposed Idea will be portrayed in the following part.

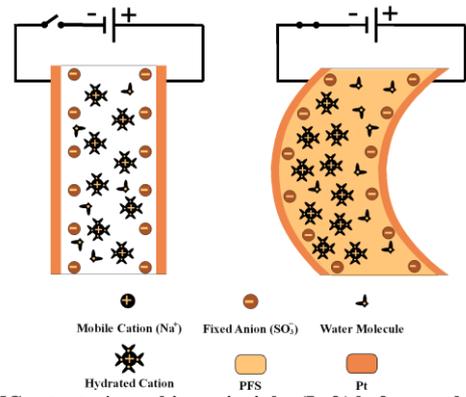

**Fig. 1. IPMC actuator's working principle: (Left) before applying voltage; (Right) after applying voltage[17].**

## III. MICROFLUIDIC MICROMIXERS

Designing micromixers is an inevitable challenge in microfluidic technology, and there are so many applications that they need to use micromixers. Microfluidic micromixers can be applied to widespread applications in Biological and Chemical processes like biomedical analysis, oligonucleotide synthesis, spectroscopic techniques and so on. For example, an emerging approach in biomedical analysis is designing enzyme-based biosensors, in which the early detection is highly dependent on the rate of an enzyme-catalyzed reaction. For example, in diabetes, glucose sensors have been commercialized by employing glucose oxidase (GOX) as their enzyme. The resulting enzyme-catalyzed reaction of glucose with oxygen produces gluconolactone and $H_2O_2$ to measure the concentration of Glucose in any individual blood. The reaction between Glucose and glucose oxidase (GOX) can be facilitated by introducing these reagents into the mixer, which will help in measuring the Glucose concentration precisely and rapidly[22]. We can find some other biological application of micromixers, for example solid-phase synthesis of oligonucleotides [23] and studying folding/unfolding structures of proteins [24] are two important applications.

One of the challenges in front of microfluidic-based devices and specially micromixers is that behavior of the fluids in microchannels of the chip differs from what is expected in a macro-sized channel. The reason behind this anomaly can be explained with the Reynolds number. The Reynolds number is a dimensionless quantity that correlates inertia forces of a fluid to its viscous forces and is defined as:

$$Re = \frac{Inertial\ Forces}{Viscous\ Forces} = \frac{\rho V \cdot \frac{d^2 x}{dt^2}}{\mu \frac{du}{dy} \cdot A} = \frac{\rho \upsilon L}{\mu} \qquad (1)$$

Where $\rho$ is the fluid density ($kg/m^3$), $V$ is the volume of a fluid element, $\mu$ is the dynamic viscosity of the fluid ($kg/m.s$), $u$ is local flow speed ($dx/dt$), A is the cross-sectional area of the flow ($m^2$), $\upsilon$ is the average velocity of the flow *(m/s)* and *L* is the characteristic length scale (m) [25]. Low Reynolds numbers imply that viscous forces are dominant. Therefore all flow irregularities are damped out, and the flow is smooth, or "laminar." Turbulent flow occurs at high Reynolds numbers where inertial forces are dominant therefor chaotic eddies, vortices and other flow instabilities are produced. The transition from laminar flow to turbulent flow occurs around $Re = 2000$. For a microfluidic channel, $L$ is very small, and the $Re$ is usually much less than 10, often less than 1. This means that the flow in a microchannel is completely laminar. Despite possible advantages of laminar flow, it has a major drawback as it prevents fluids from turbulent mixing. As a result, mixing can only be achieved through diffusive mixing which is an inherently slow process and requires a long channel to achieve sufficient mixing. To address this problem, novel and various methods have been designed and examined. Like general microfluidic devices, micromixers are described in two main categories: "active micromixers" and "passive micromixers"[26]. Passive mixers are not only easier to fabricate and work but also, they do not require a power source. This makes passive mixers very suitable for a portable device like POC systems. However, since the design cannot change after the fabrication of the device, these mixers are difficult to control externally by users. This results in difficulties to achieve optimal mixing from a device when it's used in different circumstances. Also, users cannot switch off the mixing process when needed. Active micromixers, on the contrary, have better control over the mixing process and their mixing efficiency is higher than typical passive mixers. But the fabrication of an active micromixer and its integration with the device is more troublesome. In addition, there is a limit on the use of active components in POC systems since they need to be compact, cheap, portable and easily handled [27], and it is the main reason that we want to embed the IPMC soft actuator into the microfluidic micromixers.

## IV. PROPOSED SOFT MICROMIXER

Mass transfer during the mixing process in microfluidic devices generally occurs because of two different types of mixing, first is the heterogonous mixing created by convection, $-\nabla \cdot (uc)$ in Eq.(2) shows this term. The other type is homogenous mixing caused by molecular diffusion between adjacent domains which is shown as $\nabla \cdot (D\nabla c)$ in In Eq.(2). As a result, The general form of the combined convection-diffusion equation is calculated using the following equations[28]:

$$\frac{\partial c}{\partial t} = \nabla \cdot (D\nabla c) - \nabla \cdot (uc) \qquad (2)$$

Where $c$ is the species concentration, $D$ is the diffusion coefficient, v is the velocity field that the fluid is moving with, $\nabla$ represents gradient and $\nabla \cdot$ represents divergence. In a straight microchannel, with laminar flows, since convection transports mass only tangent to the velocity, that is along streamlines, it cannot lead to mass transfer between adjacent layers of parallel fluids flow. In other words, for a laminar flow at steady state, only diffusion can allow mass transfer normal to the fluid flow. On the other hand, the molecular diffusion has an inverse relationship with the size of the fluid molecules or particles to be mixed. Hence for large biomolecules or cells, the mixing procedure is very time-consuming, and it will take hours to complete [29]. In a variety of microfluidic and LOC applications such as drug delivery, cell activation, protein folding, enzyme reaction, etc, the rapid mixing is an essential action[30]. For this reason, in such microchannels and applications we need to use an actuator to mix the fluids in order to overcome mentioned constraints. Hence, having a suitable actuator like IPMC seems fine to overcome these limits in micromixers with regard to its promising features. In this paper, we will present an IPMC based micromixer for blending in microfluidic devices which can be used without any particular types of equipment. We could find only two papers [21, 31] which reported using an IPMC based micromixers although these two works are not appropriate for the specific mixing target that we have here. For instance, in [31], the size that they have validated their work for is bigger than the conventional microfluidic devices, and their structure is not suitable for microfluidic systems. In [21] a perpendicular IPMC in a microfluidic channel is proposed which is not much of use because the length of the IPMC cannot be large enough for full bending behavior and as you can find in the experimental results of this work [21], it cannot work properly for mixing purposes. The proposed micromixer is a PDMS based microfluidic T junction which a cantilever IPMC is integrated at the starting point of it (Fig.2). Here the presented active device for causing turbulence is IPMC which by applying an appropriate voltage with proper frequency will work as a high-performance mixer.

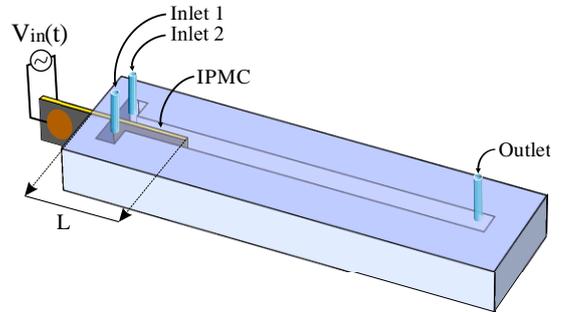

**Fig. 2. Proposed IPMC-based micromixer device.**

In the proposed micromixer, the interaction between a deformable IPMC with its surrounding fluid can change the velocity field of each flow. Therefore, by increasing the convection effect, the mixing process will be improved. In order to show this improvement, first, the velocity field of fluids should be calculated. By coupling the dynamic equation of IPMC and the Navier-stock equation, which is used to describe the dynamic of fluids, the velocity field can be calculated, $\mathbf{u} = \hat{\mathbf{u}}\omega\delta$. In other words, a fluid-structure interaction problem should be solved to find the velocity field. For a cantilever IPMC, [32] has already obtained the nondimensional forms of these coupled equations as (3) and (4) respectively.

$$\begin{cases} \hat{\nabla} \cdot \hat{\boldsymbol{u}} = 0 \\ \dfrac{1}{\beta} \dfrac{\partial \hat{\boldsymbol{u}}}{\partial \hat{t}} + (\hat{\boldsymbol{u}} \cdot \hat{\nabla})\hat{\boldsymbol{u}} = -\hat{\nabla}\hat{p} + \dfrac{1}{Re}\hat{\Delta}\hat{\boldsymbol{u}} \end{cases} \qquad (3)$$

$$\hat{w}(\hat{X},\hat{t}) = \beta \frac{W(\hat{X})}{W(1)} \sin \hat{t} \quad (4)$$

Where $\hat{\nabla}$ and $\hat{\Delta}$ are the Nabla and Laplacian operators in nondimensional spatial coordinates, $Re$ is the Reynolds number defined as $\omega \delta L / \nu$ and $\beta = \delta / L \cdot \omega$, $w(X,t)$, $\delta$, $L$ and $W(X)$ are the radian frequency, tip displacement, maximum tip displacement, length, and fundamental mode shape of IPMC where $W(X)$ is defined as follows:

$$W(X) = \sin\left(\alpha \frac{X}{L}\right) - \sinh\left(\alpha \frac{X}{L}\right) - \frac{\sin \alpha + \sinh \alpha}{\cos \alpha + \cosh \alpha}\left(\cos\left(\alpha \frac{X}{L}\right) - \cosh\left(\alpha \frac{X}{L}\right)\right) \quad (5)$$

The details of the relations and the meaning of unknown parameters and functions have been described in [32]. The viscous friction and momentum are related to Reynolds numbers. Laminar flows are normally observed in microfluidic systems due to the low Reynold number [30]. On the other hand, a turbulent flow is known by its high Reynold number. As it was at that point referenced, a turbulent transition should be made for quick mixing. Thus, an increase in the Reynold number is an essential requirement in fast mixing procedures. Extracted from coupled equations (Eqs 3 and 4), the Reynold number relate to the IPMC-fluid interaction is equal to $Re = \omega \delta L / \nu$. With regard to this equation, this number is directly related to the radian frequency ($\omega$), maximum tip displacement ($\delta$), and length ($L$) of IPMC. Hence, for quick mixing of laminar fluids and increased Reynold number, the mentioned parameters of an IPMC should be increased. In other words, IPMC can be considered as a solid stirrer that the increase in its power ( by an increase in $\delta$ ) or its speed ( by an increase in $\omega$ ) will increase the mixing rate. Hence, based on the fast mixing rate, the required perturbation of liquids in the channel can be made. The aim of this paper is to show that in case of the active component of a real-sized micromixer, an IPMC actuator can be regarded as a proper candidate. In order to investigate this capability of an IPMC, the proposed IPMC-based micromixer has been simulated using finite element approach and also it has been tested practically and, in the next part we will show that the proposed idea is working and has adequate potential to be a POC-based active micromixer.

## V. MULTIPHYSICS FEM SIMULATION

As explained in the previous section to investigate the effect of IPMC vibration on the velocity field of fluid flows the Eqs (3) & (4) should be solved fully coupled. Also, to show the effect of changing the velocity field on the mixing process, the Eq (2) needs to be solved simultaneously. The COMSOL Multiphysics FSI interface allows the coupled solving of time-dependent structural deformations and fluid flow variables in a moving mesh geometry consisting of a solid deformable object surrounded by fluids. Hence, COMSOL Multiphysics software was employed to carry out the numeric 2D simulations. The mixer was modeled using time-dependent incompressible Navier–Stokes equations in the laminar flow physics coupled with solid mechanics module that describes the dynamic behavior of deformable IPMC. Also, convection-diffusion investigate the mixing process, by adding transport of diluted species interface, the convection-diffusion equation was solved simultaneously. A non-slip FSI boundary was automatically set up along the inner wall of the channel. The boundary conditions used at the system inlets were uniform velocity perpendicular to the inlet face. The inlet flows were set as laminar flow. In order to replicate the behavior of IPMC, when an Ac voltage is applied to, a prescribed harmonic displacement was assigned to the free end of IPMC. Other parameters used in the simulation were shown in Table 1. COMSOL simulations have been conducted to show that an IPMC cantilever beam embedded in a T-junction like microchannel with real size dimensions (Fig.3) can mix two fluids appropriately. The simulations have been done for five frequencies from 1 Hz to 5 Hz IPMC tip displacements. The results of this simulations have been quantized as a mixing index. The proposed mixing index is a function of RGB euclidean distance between the average of colors in regions $r_1$ and $r_2$ (as illustrated in the end of channel in Fig.3) that this index is defined as follows :

$$Mixing\ Index(\%) = H_W \times ED_{RGB} \quad (6)$$

Where

$$ED_{RGB} = 100\left(1 - \sqrt{\frac{(R_2 - R_1)^2 + (G_2 - G_1)^2 + (B_2 - B_1)^2}{(255)^2 + (255)^2 + (255)^2}}\right) \quad (7)$$

$$H_W = \frac{1}{1 + e^{-a(|H_2 - H_1| - c)}} \quad (8)$$

**Table 1. Simulation parameters.**

| Fluid properties and boundary condition | Value |
|---|---|
| Density, $\rho$ (kg/m3) | 998 |
| Dynamic viscosity, $\mu$ (Pa.s) | $8.9 \times 10^{-4}$ |
| Molecular diffusivity, D(m$^2$/s) | $10^{-9}$ |
| Outflow pressure, P (Pa) | 0 |
| Concentrations at inlet A, c (mol/m3) | 100 |
| Concentrations at inlet B, c (mol/m3) | 0 |
| Velocity at both inlets $\mu m/s$ | 300 |

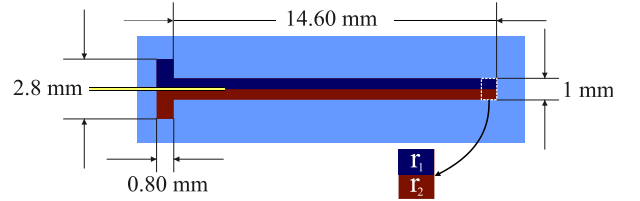

**Fig. 3.** Schematics of proposed micromixer chip with the real size.

In the mentioned equations, (7) is the euclidean distance of regions 1 and 2 ($r_1$ and $r_2$) in the *RGB* color space and (8) is a sigmoid function of Hue difference in *HSL* (Hue, Saturation, Lightness) color space, where *a* and *c* define the shape of sigmoid function. The main mixing index is the same $ED_{RGB}$ which it is corrected by $H_W$ to work properly in the whole gamut of *RGB* color space. The graphs of mixing index for 1-5 Hz have been shown in Fig.4. It is clear that IPMC can mix the input fluids in all of these frequencies. But in 1Hz it works better an its mixing index is around 95% which it comes back to this fact that the first resonance frequency of IPMC is around 1Hz and it can have higher tip displacement around this frequency. As an example the COMSOL simulation results for 1 Hz IPMC tip displacements has been depicted in Fig.5.

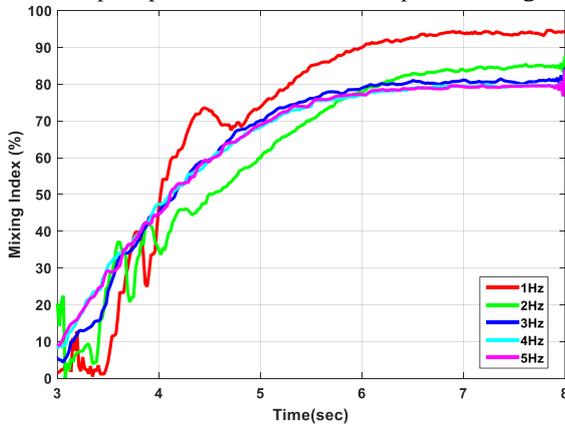

**Fig. 4. Mixing index (%) of proposed micromixer for five frequencies of 1 to 5 Hz.**

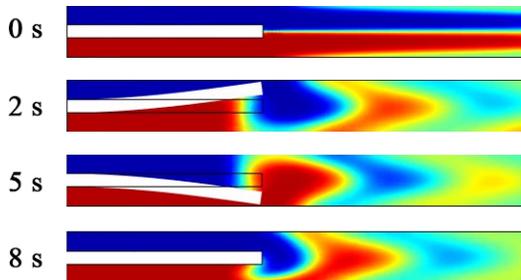

**Fig. 5. COMSOL simulation results for 1 Hz IPMC tip displacements. IPMC cantilever beam can stir two fluids together in the real size microchannel.**

## VI. FABRICATION AND EXPERIMENTAL RESULTS

In order to show the practical feasibility of the Idea, the proposed micromixer has been fabricated and tested with two dyes as input fluids. The experimental results will be discussed, but before that the fabrication procedure and required hardware for testing the chip have been described here.

### A. Chip fabrication:

As depicted in Fig.6, the procedure that we used for the fabrication of proposed microfluidic micromixer consists of a four main steps ,(1) printing the PLA mold (Fig.6-a), (2) PDMS casting (Fig.6-b), (3) oxygen plasma bonding of PDMS on the glass (Fig.6-c) and, (4) IPMC embedding into the chip (Fig.6-d). In details, we should say that, first, the chip mold was designed in a 3D sketch program (3ds max here) and printed using Polylactic acid (PLA) material in a Blackbot 3d printer. Then the primary substance for the microfluidic chip fabrication was provided by mixing the base and curing agent of PDMS ingredients (Sylgard 184) in the proportion of ten to one. After around 10 mins of mixing the parts, the mixture was full of small-sized bubbles. The mixture was placed in a vacuum chamber for 30 mins to eradicate all the bubbles from uncured PDMS. Then, the uncured PDMS was poured into the mold and placed in oven for about two hours at 85 Celsius. For the next step, the chip was placed in the oxygen plasma chamber and treated lightly by plasma for 30 seconds and immediately bonded with a clean piece of glass. Finally, the IPMC actuator was integrated into the channel. This step requires extreme expertise and precision, which is possible with experience and numerous try and failures. Because of the nature of PDMS, it is not feasible to modify or change the structure of cured PDMS, so in the process of integration, it is likely for errors to happen in coordinating the IPMC precisely in the middle of the channel. After the installation of IPMC, the gap at the peak of T-junction should get closed for channel to work correctly, but the weight of clamped IPMC caused a serious problem, and we could not fix the position of the actuator, therefore it needed to be confined, and no glue was helpful for PDMS smooth surface, so it was another challenge to finely embedding of IPMC into the channel. Finally, by utilizing the uncured PDMS as a PDMS glue and trying different efforts we could fix the position, and in order to keep the water molecule contents of IPMC we left the chip in the room temperature to be cured slowly.

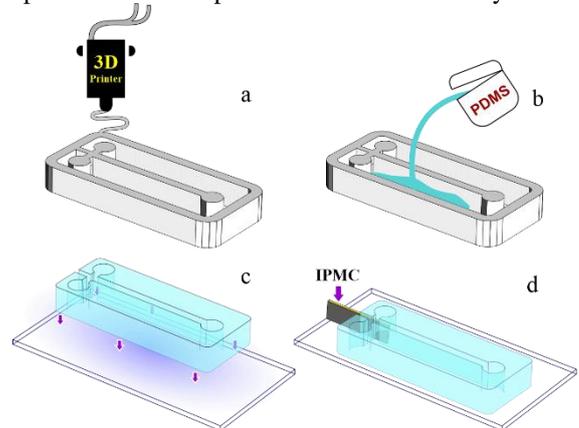

**Fig. 6. The four main steps that have been used for fabrication of the proposed micromixer, (a) printing the PLA mold, (b) PDMS casting, (c) oxygen plasma bonding of PDMS on the glass and, (d) IPMC embedding into the chip.**

### B. Hardware Apparatus

In order to generate appropriate signals for IPMC actuation, a series of different systems were put in consecutive order: At first, the desired signal was produced in a computer by MATLAB 2018 and was sent through serial port to the Data Acquisition board (DAQ). After that to amplify and shift the input signal, a differential amplifier was placed right after the DAQ board. Finally, to conduct the generated signals through IPMC, a tong holder was modified with two pieces of PCB board and linked IPMC to the amplifier output. It can be seen in Fig.7 that for capturing the movement of IPMC and fluids behavior, a precise and adjustable holder was used to place the camera on a fixed position.

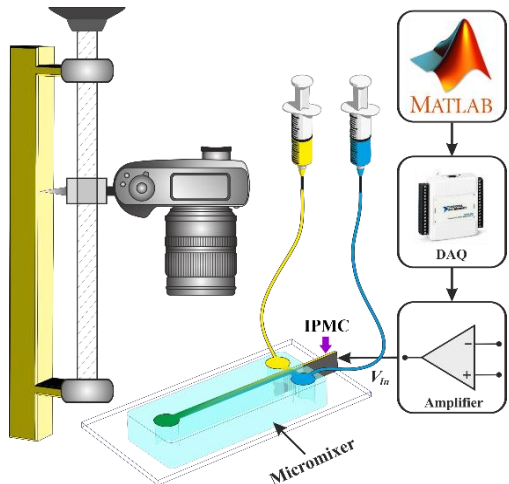

Fig. 7. The required experimental setup.

*C. Experimental Results*

To investigate the performance of the proposed micromixer, the chip was fabricated by the procedure of VI-A (Fig.8-b) and then a piece of IPMC (Fig.8-a), fabricated using the fabrication procedure of [33] was embedded into it. After that, the fabricated micromixer was tested by the mentioned setup of VI-B. As shown in Fig.7 using two syringes, two dyes are injected into the chip, and by applying a ± 3.5 Volt / 1 Hz voltage to the IPMC a perturbation is made into the channel and mixes the laminar fluids gradually. The results of this test have been shown in Fig.8-c, and as you can see here at the beginning the fluids are completely laminar, but during the time in response to IPMC fluctuation the fluids are mixed. The results show that after around 15 s the fluids are almost mixed (Fig.8-c) that it means that the real micromixer have almost twice time slower mixing speed than the simulated one. The other point that we can see in the experimental results is that the IPMC can mix the fluids and also keeps the mixed state for a long time. As it is depicted in Fig.8-d after more than 6 minutes (370 s) the fluids are still mixed. Totally the IPMCs are not mature enough *i*-EAPs, and they need some modifications to be reliable actuators. For example, the reported slowness returns back to weak mechanical properties of IPMC means that it is not strong enough to mix the fluids without attenuations in its bending magnitude. All in all, the experimental results shown that the *i*-EAPs and especially IPMCs are appropriate candidates to be fully embedded soft actuators in microfluidics devices and by improving their behavior have promising features for the next generation of electromechanical active microfluidic chips.

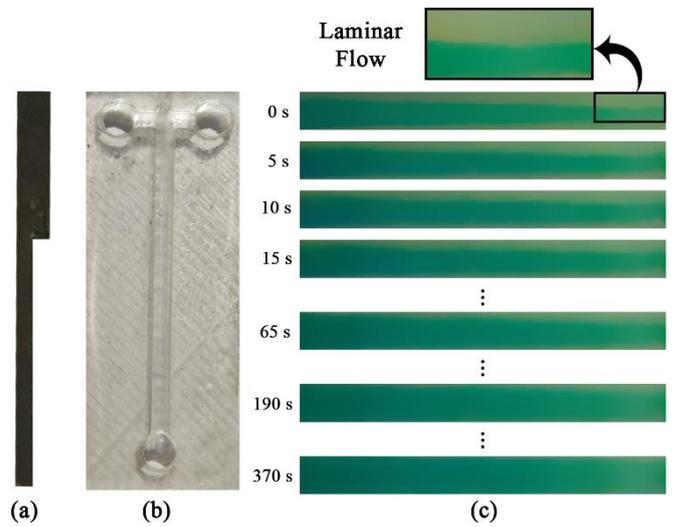

Fig. 8. (a) The used IPMC, (b) fabricated microfluidic chip, (c) experimental results of proposed micromixer.

## VII. Conclusion

POCTs help the patients to be cured at or near the bedside and far from the care centers. Besides, some POCT systems are able to send the test data to physicians or care provider centers and also, can save the data into the big databases. Hence, physicians can serve remotely and virtually at the proper time, and also some emerging technologies like Internet of Things (IoTs), deep learning, etc. can provide a new generation of health care services. For example, a smart POCT system using a deep learning algorithm can predict an urgent symptom of a disease and an IoT system can aware the nearest hospital to presents its urgent cares at the point of care. Hence research around POCT technologies is unavoidable fact. Microfluidics is one of the pioneer fields that have significant contribution in POCT systems. Active microfluidic devices are more functional but due to their needed accessories are less efficient for POCT-based applications. To find a solution to this problem, we proposed IPMC soft actuator as a new microfluidic active component that needs less and ubiquitous facilities. In the direction of this aim and, as a case study, a new type of active micromixers using IPMC soft actuator was designed and fabricated. In order to show the performance of the proposed micromixer, it was tested practically and simulated in COMSOL Multiphysics as well and the results confirmed the expected potential of IPMCs as active components in microfluidic micromixer.

### Acknowledgment

This work has been supported by Iran Biotechnology Council [11/26236]